\documentclass[a4paper, 10pt, conference]{article}      

\usepackage{graphics} 
\usepackage{epsfig} 
\usepackage{amsmath} 
\usepackage{amssymb}  
\usepackage{algorithm}
\usepackage{algpseudocode}%

\newtheorem{remark}{Remark}

\providecommand{\matlab}[0]{\textsc{Matlab}~}

\providecommand{\lpv}[0]{\textbf{LPV}~}
\providecommand{\lfr}[0]{\textbf{LFR}~}

\providecommand{\ie}[0]{\emph{i.e.}~}

\providecommand{\eg}[0]{\emph{e.g.}~}

\newenvironment{eq}{\everymath {\displaystyle \everymath{ }} \equation}{ \endequation} %
\providecommand{\norm}[1]{\left\lVert #1 \right\rVert} %
\providecommand{\x}[0]{\mathbf{x}} %
\providecommand{\dx}[0]{\mathbf{\dot{x}}} %
\renewcommand{\u}{\mathbf{u}} %
\providecommand{\y}[0]{\mathbf{y}} %
\providecommand{\pv}[0]{\mathbf{p}} %

\providecommand{\Hreal}[0]{\mathcal{S}} %
\providecommand{\Htran}[0]{\mathbf{H}} %
\providecommand{\A}[0]{{A}} %
\providecommand{\B}[0]{{B}} %
\providecommand{\C}[0]{{C}} %
\providecommand{\D}[0]{{D}} %

\providecommand{\Htwo}[0]{{\mathcal{H}_{2}}} %
\providecommand{\Hinf}[0]{{\mathcal{H}_{\infty}}} %
\providecommand{\Cplx}[0]{\mathbb{C}} %
\providecommand{\Real}[0]{\mathbb{R}} %
\providecommand{\matrixthree}[9]{ \left[\begin{array}{ccc} #1 & #2 & #3 \\ #4 & #5 & #6 \\ #7 & #8 & #9\end{array}\right] } %
\title{Structured linear fractional parametric controller \\ design with $\Hinf$ performances}

\author{C. Poussot-Vassal
\thanks{Onera - The French Aerospace Lab, F-31055 Toulouse, France; email: {charles.poussot-vassal@onera.fr}}%
}

\begin{document}

\maketitle
\thispagestyle{empty}
\pagestyle{empty}

\begin{abstract}
This paper proposes an simple but yet effective approach to structured parametric controller design in a linear fractional form. The main contribution consists in using structured $\Hinf$ oriented optimization tools in an original manner to either \emph{(i)} construct a parametric controller or \emph{(ii)} a family of controllers with varying performances. Practical and numerical issues are also discussed to provide readers and practitioners a simple way to deploy the proposed process. The overall approach is illustrated through two numerical academical (but still complex) examples illustrating two applications: first, a parametric controller design adapted to a parameter dependent model of a clamped beam and, second, a controller with parameter dependent performance applied on a building model.
\end{abstract}

\section{Introduction}

\subsection{Motivating context and problem formulation}

In numerous industrial applications, the $n$-th order $n_u$ inputs $n_y$ outputs linear dynamical model describing a system can either be given in an invariant form as $\Htran(s)=\C(sI_n-\A)^{-1}\B+\D \in \mathcal H_\infty^{n_y \times n_u}$, equipped with realization $\Hreal:(\A,\B,\C,\D)$ defined as
\begin{eq}
\dx(t) = \A\x(t)+\B\u(t), \y(t) = \C\x(t)+\D\u(t), 
\label{eq-lti}
\end{eq}
or in a parametric form  $\Htran(s,\pv)=\C(\pv)\big(sI_n-\A(\pv)\big)^{-1}\B(\pv)+\D(\pv) \in \mathcal H_\infty^{n_y \times n_u}$, equipped with realization $\Hreal(\pv):\big(\A(\pv),\B(\pv),\C(\pv),\D(\pv)\big)$ defined as
\begin{eq}
\dx(t) = \A(\pv)\x(t)+\B(\pv)\u(t), \y(t) = \C(\pv)\x(t)+\D(\pv)\u(t), 
\label{eq-lp}
\end{eq}
where $\x(t)\in\Real^{n}$, $\u(t)\in\Real^{n_u}$, $\y(t)\in\Real^{n_y}$ and $\pv\in\mathcal P \subseteq\Real^{n_p}$ represent the state, input, output and parameter vectors, respectively. Moreover, the $\mathcal P$ subspace is closed, the Laplace variable is denoted $s$ and the $\A$, $\A(\pv)$, $\B$, $\B(\pv)$, $\C$, $\C(\pv)$, $\D$ and $\D(\pv)$ matrices are of appropriate dimension\footnote{Throughout this paper, we denote $\mathcal{H}_2^{n_y\times n_u}$ (resp. $\mathcal{H}_\infty^{n_y\times n_u}$) or simply $\mathcal{H}_2$ (resp. $\mathcal{H}_\infty$), the open subspace of $\mathcal{L}_2$ (resp. $\mathcal{L}_\infty$) with matrix-valued function $\Htran(s)$ with $n_y$ outputs, $n_u$ inputs, $\forall s \in \Cplx$, which are analytic in $\textbf{Re}(s)> 0$ (resp. $\textbf{Re}(s)\geq 0$)}. 

\begin{remark}[Parametric vs. \lpv]
It is noteworthy to distinguish the \textbf{parametric} form with the Linear Parameter Varying (\textbf{LPV}) one. Indeed while in the latter case the parameter $\pv$ is considered as varying, in the former one (which we consider in this paper), the parameter can simply be a frozen physical system coefficient (\eg the geometrical parameters of an aircraft wing \cite{Gadient:2012}, or the section of an open-channel \cite{DalmasECC:2016}). Parametric models can appear when the system's model is $\pv$ dependent \cite{VuilleminCST:2017} or when the $\pv$ parameter is an artificial one that characterizes the closed-loop performances (see the section examples).
\end{remark}

In the parametric case $\Htran(s,\pv)$ \eqref{eq-lp}, it is interesting being able constructing a controller, function of the parameter $\pv$ value, that reaches a given performance level. Similarly, in the invariant case $\Htran(s)$ \eqref{eq-lti}, for practical reason, one can be interested in constructing a family of (parametric) $\pv$ dependent controllers, achieving varying performances, which can be tested and directly adjusted on the real system during tests validations\footnote{This last case is particularly interesting when real tests are costly and engineers cannot stop the process, re-tune the law and re-start the tests. An illustration of this situation can be found in aeronautics, \eg for aircraft flight and ground tests, as in \cite{MeyerMOVIC:2016}.}. Parametric controller design is then clearly a challenging task for many industrial applications since it provides the possibility to tune the control performance according to the plant configuration or to provide practitioners the ability to test a family of control laws in a simpler manner.  

More specifically, in the considered framework, given a model as in \eqref{eq-lti} or \eqref{eq-lp}, we aim at synthesizing a $n_K$-th order $\pv$ dependent controller $\mathbf K^\star(s,\pv)\in \mathcal H_\infty^{n_u \times n_y}$ described as in the following a Linear Fractional (\textbf{LF}) structure:
\begin{eq}
\mathbf K^\star(s,\pv) = \mathcal F_u\big(\mathbf K(s),\Delta\big),
\label{eq:kobj}
\end{eq}
that ensures closed-loop stability and achieve some $\Hinf$ performances (the performance is more precisely defined later). Let $\mathbf K(s)\in \mathcal K \subseteq \mathcal H_\infty^{n_u\times n_y}$, $\Delta = \pv I_{n_\Delta}\in \Real^{n_\Delta \times n_\Delta}$ and $\mathcal F_u(.)$ denotes the upper linear fractional operator \cite{MagniLFR:2006}, defined as (for appropriate partitions of $M$ and $\Delta$) by
${\cal F}_u(M,\Delta) =  M_{22}+M_{21}\Delta(I-M_{11})^{-1}M_{12}$.

Obviously, many solutions have been derived in the literature to design such a parametric controller \eqref{eq:kobj} (with varying structure). Among the methods, the so-called \lpv community did provide a lot of very interesting tools and procedures (still, mostly oriented to varying parameters, see \eg \cite{Wang:2016}). Moreover, the robust control community also introduced a set of mathematical results in this sense such as the Linear Fractional Representation (\textbf{LFR}) framework and the associated control set-up (see \eg the approaches addressing the  $\Hinf$ norm \cite{Gahinet:1994,Henrion:2004,HIFOO:2006,Apkarian:2006} or the $\Htwo$ one \cite{deSouza:2005}). Moreover, the Youla parametrization also provides a framework to this aim (see \eg \cite{Neering:2006} for more details). For additional information on theses subjects, reader is invited to refer to the many results of \eg C. Scherer \cite{Scherer:1996,Scherer:1997}, P. Apkarian \cite{Apkarian:1995,Gahinet:1996,Apkarian:2006}, G. Balas \cite{Fialho:2002,Hjartarson:2013} and co-workers.

\subsection{Contributions and outlines}

The result provided in this paper aims at addressing the problem of structured parametric controller design \eqref{eq:kobj} in the linear framework for \eqref{eq-lti} and \eqref{eq-lp} models. More specifically, a simple but yet very effective methodology to design such $\pv$ dependent controller (or controller family)  achieving $\Hinf$ performances, is detailed in the rest of the paper. In addition, using the recently developed structured $\Hinf$ oriented optimization tools made available in \matlab through the \texttt{hinfstruct} method \cite{Apkarian:2006}, we also provide a detailed approach with numerical issues to deal with this problem in order to  given practitioners keys to solve this kind of problem.

The remaining of the paper is organized as follows: the main result, \ie the synthesis of a structured  parametric linear fractional controller achieving $\Hinf$ performances is detailed in Section \ref{section-result}. Then, Section \ref{section-numerical}, provides practitioners some numerical and practical issues to easily optimize such a controller. Numerical examples are given in Section \ref{section-examples} detailing the design of a structured parametric controller in a linear fractional form for two interesting cases: first, based on a parametric model of a clamped beam, and secondly, based on a non-parametric model of a building, but including parametric closed-loop performances. Discussions close the paper in Section \ref{section-conclusions}.

\section{Main result: structured linear fractional parametric controller synthesis}
\label{section-result}
\subsection{Problem formulation with $\Hinf$ performances}

Let us consider a linear dynamical model of the form \eqref{eq-lti} or \eqref{eq-lp}. As evoked in the introductory part, we aim at designing a $\pv$ dependent parametric controller in linear fractional form that ensures some $\Hinf$ performances. As it is standard in the robust framework, let us first define the following generalized plant $\mathbf{T}(\pv)=\mathbf W_i(s) \Htran(s,\pv) \mathbf W_o(s,\pv)$, where, $\mathbf W_i(s)$ and $\mathbf W_o(s,\pv)$ are the weighting filters defying the input and parametric (or not) output signals. Both $\mathbf W_i(s)$ and $\mathbf W_o(s,\pv)$ are constructed by the user to define the desired performances attenuation and its bandwidth. The associated state-space realization is then given by\footnote{Note that according to the original plant, the parameter dependency can either come from the system model itself, if described by \eqref{eq-lp}, or from the performance weighting filters $\mathbf W_o(s,\pv)$.}, 
\begin{eq}
\left\{
\begin{array}{rcl}
\dx(t) &=& \A(\pv) \x(t) + \B_1(\pv) \mathbf w(t) + \B_2(\pv) \u(t)  \\
\mathbf z(t)&=& \C_1(\pv) \x(t) + \D_{11}(\pv) \mathbf w(t) + \D_{12}(\pv) \u(t)\\
\y(t)&=& \C_2(\pv)\x(t) + \D_{21}(\pv)\mathbf w(t) + \D_{22}(\pv) \u(t)
\end{array} \right. 
\label{eq:gene}
\end{eq}
where $\x(t) \in \Real^{n}$, $\mathbf w(t) \in \Real^{n_w}$, $\u(t) \in \Real^{n_u}$, $\mathbf z(t) \in \Real^{n_z}$ and $\y(t) \in \Real^{n_y}$ are the states, exogenous input, single control input, performance output and measurement signals, respectively. Then, the associated performance transfer from $\mathbf w(t)$ to $\mathbf z(t)$, parametrized by $\pv$, is defined as,
\begin{eq}
\mathbf T(s,\pv) =\mathbf W_i(s) \Htran(s,\pv) \mathbf W_o(s,\pv).
\end{eq}
Then, mathematically, the $\Hinf$ parametric control design objective consists in finding the optimal controller $\mathbf K^\star(s,\pv)$ such that,
\begin{eq}
\small
\mathbf K^\star(s,\pv) := \arg \min_{\mathbf{K}\in\mathcal K} \max_{\pv\in\mathcal{D}} \norm{\mathcal F_l\Big(\mathbf{T}(s,\pv),\mathcal F_u\big(\mathbf{K}(s),\Delta\big)\Big)}_{\mathcal H_\infty}
\label{eq:pb}
\normalsize
\end{eq}
where $\mathcal F_l(.)$ denotes the lower linear fractional operator defined as ${\cal F}_l(P,\Delta) = P_{11}+P_{12}\Delta(I-P_{22})^{-1}P_{21}$, $\mathbf{T}(s,\pv)\in\mathcal{H}_\infty$ is the parameter dependent generalized plant performance transfer, $\Delta \in \mathbb{R}^{n_\Delta\times n_\Delta}$ is a user-defined diagonal structure gathering the parametric variation of $\pv \in \mathcal{P} \subseteq \mathbb{R}^{n_p}$ ($\mathcal{P}$ is a closed set). Finally, $\mathbf K(s)\in \mathcal K\subseteq\mathcal{H}_\infty$ is the controller (dynamical operator) to be found. This last dynamical system might then be structured as \emph{(i)} a full block or \emph{(ii)} a sparse matrix, affine or not (see next Section \ref{subsection-K} for details on how to structure such operator),

\subsection{Solution as a linear fractional form}

We aim at solving \eqref{eq:pb} with a controller in a linear fractional form. Consequently the, controller description can be recast as:
\begin{eq}
\mathcal F_u\big(\mathbf{K}(s),\Delta\big) := \mathcal F_u\Big( \mathcal F_u\big(K,\frac{1}{s}I_{n_K}\big),\Delta\Big),
\label{eq:Klp}
\end{eq}
where,
\begin{eq}
K = \matrixthree{A_K}{B_w}{B_u}{C_z}{D_{zw}}{D_{zu}}{C_y}{D_{yw}}{D_{yu}} \in \Real^{n_K n_\Delta n_u \times n_K n_\Delta n_y}.
\label{eq:Kmatrix}
\end{eq}
Then, problem \eqref{eq:pb} turns to seek for $K^\star$ such that it solves the problem given in \eqref{eq:pb2}.

It is now clear that if a solution of \eqref{eq:pb2} is found, then one can reconstruct the parametric controller $\mathbf K^\star(s,\pv)$ through relation \eqref{eq:Klp}. Nevertheless, \eqref{eq:pb2} still requires to solve for all $\pv \in \mathcal P$, which leads in practice to a infinite number of $\Hinf$ problems to compute. As in the existing robust control framework, an alternative consists in solving the above problem for a finite number $M \in \mathbb N$ of evaluation $\pv_j$ ($j=1,\dots,M$) values of $\pv$, as exposed in \eqref{eq:pb3}.
\begin{figure*}
\begin{eq}
K^\star := \arg \min_{K\in \Real^{n_K n_\Delta n_u \times n_K n_\Delta n_y}} \max_{\pv\in\mathcal{D}} \norm{\mathcal F_l\bigg(\mathbf{T}(s,\pv),\mathcal F_u\Big( \mathcal F_u\big(K,\frac{1}{s}I_{n_K}\big),\Delta\Big)\bigg)}_{\mathcal H_\infty}
\label{eq:pb2}
\end{eq}
\begin{eq}
K^\star := \arg \min_{K\in \Real^{n_K n_\Delta n_u \times n_K n_\Delta n_y}} \max_{\pv_j\in\mathbb{R} ~ j=1,\dots,M} \norm{\mathcal F_l\bigg(\mathbf{T}(s,\pv_j),\mathcal F_u\Big( \mathcal F_u\big(K,\frac{1}{s}I_{n_K}\big),\Delta_j\Big)\bigg)}_{\mathcal H_\infty}
\label{eq:pb3}
\end{eq}
\end{figure*}

If a solution of \eqref{eq:pb3} is found, then, following \eqref{eq:Klp}, the optimal parametric controller $\mathbf K^\star(s,\pv)$ in linear fractional form is obtained. Its implementation is then straightforwardly done on any computer-aided system. Before entering  into numerical considerations, that are crucial for successful application, let us now be more specific on the controller subset $\mathcal K$, defining the controller structure and on the implications on the matrix $K$ \eqref{eq:Kmatrix}.

\subsection{Considerations about $\mathcal K$ and parameter dependency}
\label{subsection-K}

Now the theoretical problem have been set-up and, let us provide some insight on the parametrization of the solution through the $\mathcal K$ subspace, the $\Delta$ block and  especially the $K$ matrix \eqref{eq:Kmatrix}.

\subsubsection{The $\mathcal K$ subspace and $K$ structure}

with reference to the original optimisation problem \eqref{eq:pb}, the decision variable is $\mathbf K(s)$ which belongs to ${\mathcal K}$. Let us derive some specific user cases of sets ${\mathcal K}$ and its implication on the problem solved in \eqref{eq:pb3}:
\begin{itemize}
\item if ${\mathcal K}=\mathcal H_\infty^{(n_u\times n_y)}$, then, $K$ \eqref{eq:Kmatrix} is a full block matrix. This stands as the most generic case where all variables in $K$ are adjustable and the controller obtained might be proper and the dependency with $\pv$ is rational. 
\item if ${\mathcal K}=\mathcal H_2^{(n_u\times n_y)}$, then, $K$ \eqref{eq:Kmatrix} is a full block matrix except for the $D_{yu}$ term and $D_{zu}$ and/or $D_{yw}$ which are null. The controller obtained controller is strictly proper (rolls-off in high frequencies) and the dependency with $\pv$ might be  rational too. 
\end{itemize} 

\subsubsection{The parametric dependency (rational vs. affine) and $K$ structure}

in addition, with reference to \eqref{eq:Kmatrix}, let us recall the controller realization at $\pv_j$, associated with the linear fraction operators as
\begin{eq}
\begin{array}{rl}
\mathbf K(\Delta_j) :=& \big(A_K + B_w\Delta_jM_jC_z,B_u + B_w\Delta_jM_jD_{zu}, \\
&C_y + D_{yw}\Delta_j M_j C_z, D_{yu} + D_{yw}\Delta_jM_jD_{zu}\big) \nonumber
\end{array}
\end{eq}
where $M_j=(I_{n_\Delta}-D_{zw}\Delta_j)^{-1}$.
Then, 
\begin{itemize}
\item if $D_{zw}\neq 0$, $\mathbf K(s,\pv)$ parameter dependency is rational,
\item if $D_{zw}= 0$, $\mathbf K(s,\pv)$ parameter dependency is affine.
\end{itemize} 
We will see in the example that this selection can impact the solution. Still reader should keep in mind that the rational case theoretically provides a less conservative solution than the affine one. However, this conservative property is balanced by a more complex parametrization, which in practice can be a brake in the optimization procedure (see later).

\subsubsection{The parametric dependency order $n_{\Delta}$}

in problem \eqref{eq:pb3}, we consider the $\Delta$ block being known. Obviously, in the complete problem, this variable is not known by advance and additional research should be done to consider it as a tuning variable, and this is not the scope of this work. However, it is known from the \lfr community that such block is diagonal and thus provides a repetition of the parameters $\pv$. In this preliminary study we simply focus on the case where $\Delta=\pv I_{n_\Delta} \in \Real^{n_{\Delta}\times n_{\Delta}}$. Then, the only tuning variable that a user has to deal with is the dimension $n_{\Delta}$. So far, as illustrated later in the examples, no clear solution on the mechanism to implement is known and the optimal choice is dependent on both the complexity and representativeness of this structure. However, it is to be kept in mind that a $n_{\Delta}=0$ implies a non parametric controller (\ie classical controller) and $n_{\Delta}>0$ lead to parametric control with an enhanced performance with accompanied with an increasing complexity.

\section{Numerical issues discussion}
\label{section-numerical}

\subsection{Problem parametrization}

As detailed above, the considered optimization problem \eqref{eq:pb3} is then function of $K \in \Real^{n_K n_\Delta n_u \times n_K n_\Delta n_y}$ and $M\in \mathbb N$. The optimization problem parametrization contains $n_K^2 n_\Delta^2 n_u n_y$ real variables to solve. Obviously, $n_u$, $n_y$ are imposed by the sensors and actuators set-up of the control problem, one still has to choose $n_\Delta$ (the parametric complexity) and $n_K$ (the controller order). While problem of  $n_\Delta$ has already been evoked in the above section, the remaining coefficient to deal with is the order $n_K$ of the controller. This later is generally taken low to achieve a reduced order controller. However even with a low $n_K$, the parameter number increase in a square way. Consequently it is consistent to simplify the number of parameters. With reference to \eqref{eq:Kmatrix}, one way to deal with this, is to consider the following tridiagonal $A_K$ matrix structure:
\begin{eq}
A_K =\left( \begin{array}{ccccccc} 
\times&\times&0&\dots & &\dots&0\\ 
\times&\times&\times&0&& \dots&0\\ 
0&\times&\times&\times&0&\dots&0 \\
\vdots&0&\ddots&\ddots&\ddots&\ddots&0 \\
0&\dots&0&\ddots & &\times&0 \\
0&\dots&& 0&\times &\times&\times \\
0&\dots&&&0& \times&\times
\end{array}\right).
\end{eq}
Representing the $A_K$ matrix as it, is non conservative in theory and provides a considerable few variable to tune. It has show an effectiveness in many numerical applications (from model identification, approximation and control).

\subsection{Initialization of the optimization problem}

Still, it is well known that problem \eqref{eq:pb3} is NP complex and no global optimal solution can be guaranteed \cite{Apkarian:2006,HIFOO:2006}. Therefore, from a practical point of view, the initialization phase play a crucial role. This why, after having parametrized the above problem, the author advice any user to initialize the problem of parametric controller design $K$ \eqref{eq:Kmatrix} as follows: set $A_K$, $B_u$, $C_y$  and $D_{yu}$ to gain values obtained with \eg an un-parametrized $\Hinf$ optimization problem solved at the nominal $\pv$ value. Then, set $B_w$, $C_z$, $D_{zw}$, $D_{zu}$ and $D_{yw}$ to null value.

\subsection{Parametric controller stability issue}

As a practical remark, one might look for a parametric controller which is stable. For safety reasons, this kind of requirement is often requested in applications such as aeronautics, hydraulic systems. to address this constrain, following the linear fractional formulation of the problem, in addition to the problem \eqref{eq:pb3}, the following constrain can also be solved simultaneously: 
\begin{eq}
\norm{W_K\mathcal F_u\Big( \mathcal F_u\big(K,\frac{1}{s}I_{n_K}\big),\Delta\Big) }_{\mathcal H_\infty} < \gamma
\label{eq:stab}
\end{eq}
where $W_K\in\mathcal H_\infty$ is a weighting function and $\gamma$ is the performance objective of the $\Hinf$ control problem. 

\subsection{A \matlab based solution}

To give a more practical insight of the practical solution, let us provide a few lines of \matlab code to implement an such approach in a very simple way (note that here, a full block $K$ structure is considered). Let first define the integral operator and $K\in \Real^{n_K n_\Delta n_u \times n_K n_\Delta n_y}$ matrix as :\\
\texttt{Is    = tf(1,[1 0]);}\\
\texttt{K0 = [A Bw Bu; Cz Dzw Dzu; Cy Dyw Dyu];}\\
\texttt{K  = realp('K',K0);}\\
Then, by considering the generalized plants $\mathbf{T}(s,\pv_j)=$\texttt{T\{j\}} (for \texttt{j=1,...,M}) already set-up, construct the optimization problem \eqref{eq:pb3} as (where $n_K=$\texttt{nk}, $n_\Delta=$\texttt{ndelt}) \\
\texttt{Ttot = [];}\\
\texttt{for j = 1:M}\\
\texttt{Deltaj = eye(ndelt)*p(j);}\\
\texttt{Klfj  = lft(Deltaj,lft(Is*eye(nk,nk),K));}\\
\texttt{Ttot = append(Ttot,lft(T\{j\},Klfj));}\\
\texttt{Ttot = append(Ttot,Klfj*Wk);}\\
\texttt{end}\\
With reference to the loop of the above code, the first line corresponds to the evaluation of the $\Delta$ block at $\pv_j$, the second, to the evaluation of $\mathbf K(s,\pv_j)$ and the third/fourth to the concatenation of the $\mathcal F_l(\mathbf{T}(s,\pv_j),\mathcal F_u( \mathcal F_u\big(K,\frac{1}{s}I_{n_K}),\Delta_j))$ in the structure \texttt{Ttot}. Note also that \texttt{Klfj*Wk} is added to ensure controller stability and a given roll off dictated by $W_K$, as in \eqref{eq:stab}. Finally, the above problem is solved through the \texttt{hinfstruct} function as:\\
\texttt{[Kopt,gamma,info] = hinfstruct(Ttot);} \\
This leads to the $K$ matrix which cans then be used to construct $\mathbf K(s,\pv)$ easily.

\section{Numerical examples}
\label{section-examples}

Let us now, based on two academic examples, illustrate the efficiency of the proposed approach. To this aim, two use-cases are considered. The first one is a parametric model representing a clamped beam which length $L=\pv$ is the model parameter. Obviously, this parameter is not time / state dependent but simply a geometrical parameter of the system. The second use-case is a linear model of a building for which parametric performances are reached. 

\subsection{Clamped beam parametric model}

The first considered example is the Timoshenko clamped beam, which model is made available in \cite{PanzerBeam:2009}. This model is a single input single output model ($n_u=1$, $n_y=1$), and its dynamical matrices are obtained by finite element meshing. In the considered case, we select a meshing of $6$ nodes. The resulting model is then of dimension $n=60$. In addition, one interesting point of this model is that it is parametrizable with the length $L$ of the beam. In our case we consider this length varying between 10 and 20m. In this case, it means that the model \eqref{eq-lp} is available and one obtains $\Htran(s,L)=\C\big(sI_n-\A(L)\big)^{-1}\B+\D \in \mathcal H_\infty^{1 \times 1}$. The objective considered in this case is to minimize the $\Hinf$ norm of the only input/output transfer (the extremity vertical force to the vertical displacement) with a parametric controller $\mathbf K^\star(s,L)$. More specifically, the following generalized plant is considered: 
\begin{eq}
\left\{
\begin{array}{rcl}
\dx(t) &=& \A(L) \x(t) + \B \mathbf w(t) + \B \u(t)  \\
\mathbf z(t)&=& \C \x(t) \\
\y(t)&=& \C\x(t)  
\end{array} \right.  .
\label{eq:geneCBM}
\end{eq}

\begin{figure*}
\centering
\includegraphics[width=\columnwidth]{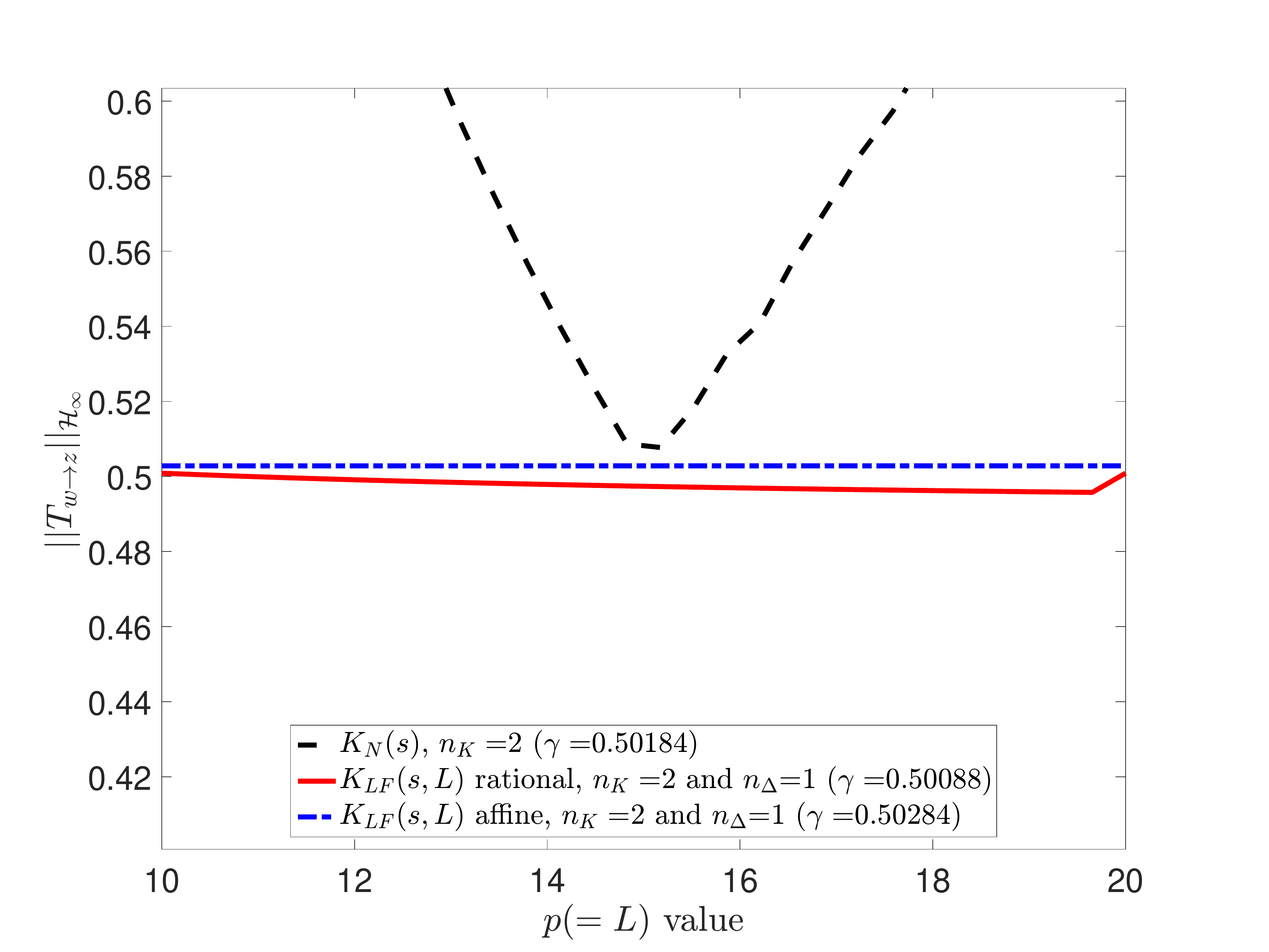}
\includegraphics[width=\columnwidth]{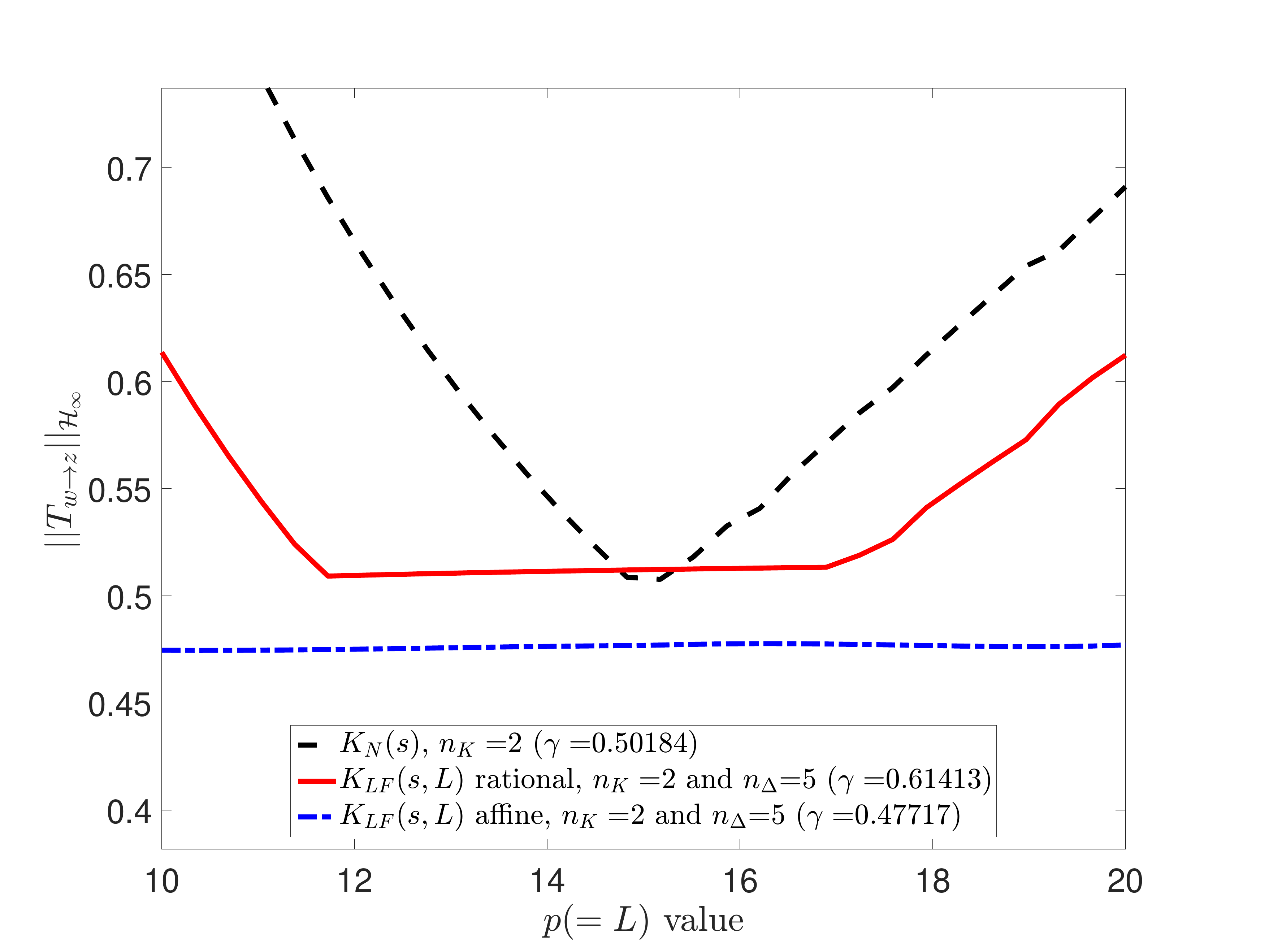}
\caption{Left frame $n_\Delta=1$ and $n_K=2$, right frame $n_\Delta=5$ and $n_K=2$. $\Hinf$ performances for varying length $L$ values of the Timoshenko clamped beam. With $\mathbf K_{N}(s)$, a non parametric controller synthesized on the nominal case ($L=15$ black dashed), $\mathbf K_{LF}(s,L)$ rational (resp. $\mathbf K_{LF}(s,L)$ affine) a parametric and rational (resp. affine) controller synthesized using $M$ configurations (red solid, resp. blue dash dotted).}
\label{fig:CBM_affine_nc2}
\end{figure*}

\begin{figure*}
\centering
\includegraphics[width=\columnwidth]{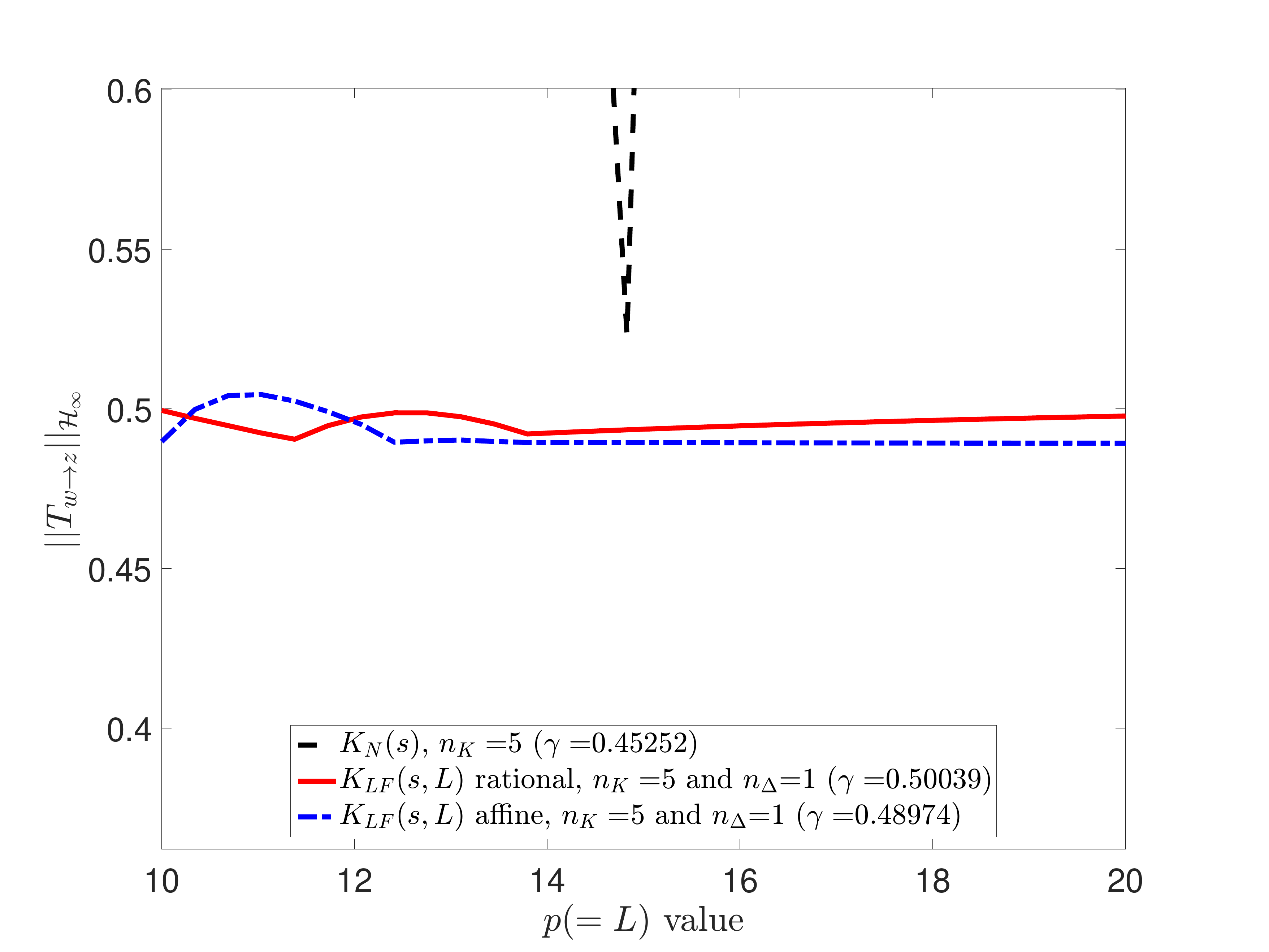}
\includegraphics[width=\columnwidth]{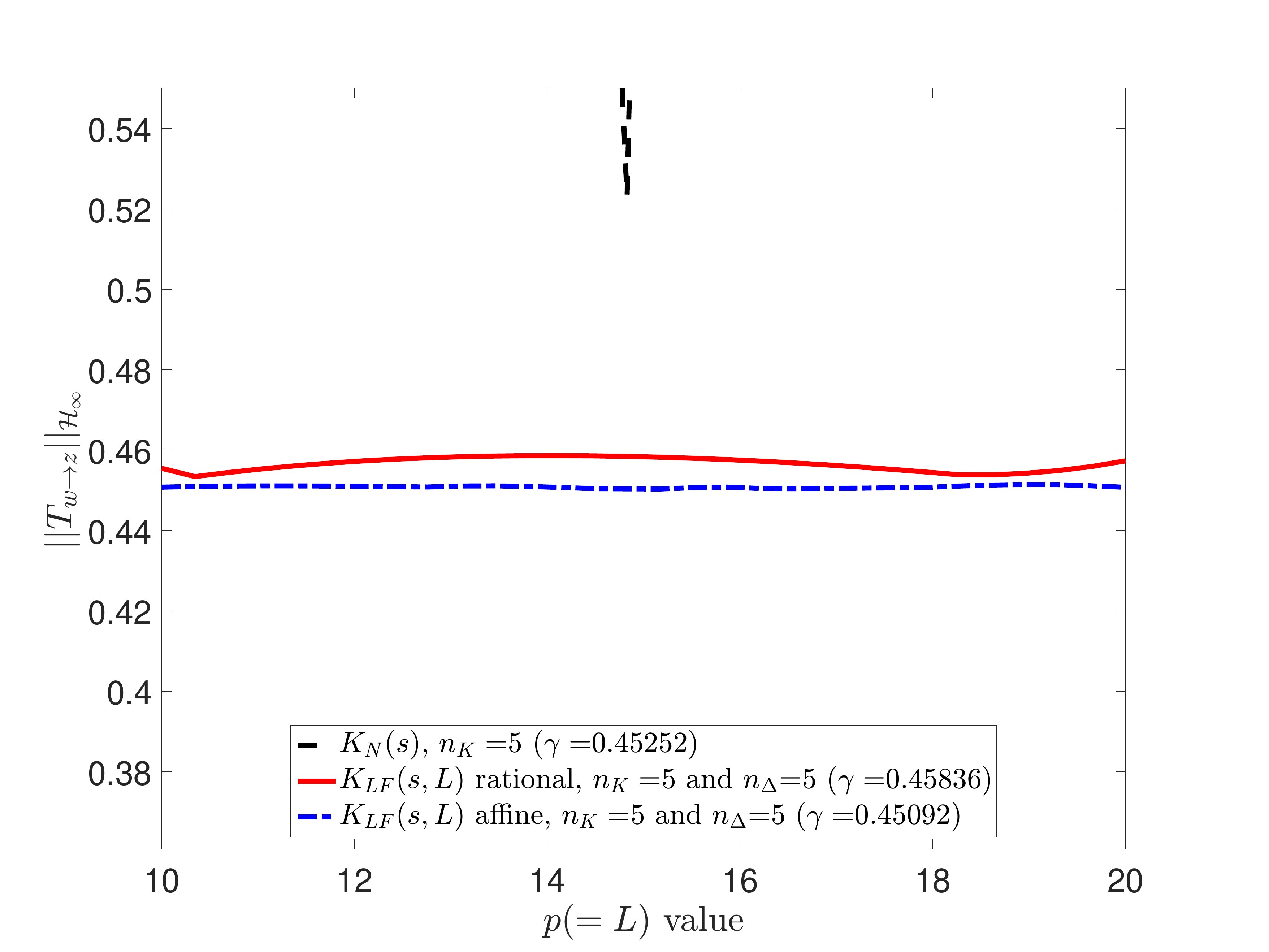}
\caption{Left frame $n_\Delta=1$ and $n_K=5$, right frame $n_\Delta=5$ and $n_K=2$. $\Hinf$ performances for varying length $L$ values of the Timoshenko clamped beam. With $\mathbf K_{N}(s)$, a non parametric controller synthesized on the nominal case ($L=15$ black dashed), $\mathbf K_{LF}(s,L)$ rational (resp. $\mathbf K_{LF}(s,L)$ affine) a parametric and rational (resp. affine) controller synthesized using $M$ configurations (red solid, resp. blue dash dotted).}
\label{fig:CBM_affine_nc5}
\end{figure*}

To be complete, a stability (and bandwidth) constrain is also added to the problem with $W_K=\frac{10^{-1}}{s/100+1}$, in \eqref{eq:stab}. Then, the procedure exposed in Section \ref{section-result} is applied for $L=\{10,12.5,15,17.5,20\}$ (\ie $M=5$) and for different $n_K$ and $n_\Delta$ values. Then, the $\Hinf$ norm of the single input single output transfer $\mathbf T(s,L)$ is evaluated for varying frozen values of $L\in[10~20]$. Some results are reported on Figures \ref{fig:CBM_affine_nc2} and \ref{fig:CBM_affine_nc5}. 


With reference to Figures  \ref{fig:CBM_affine_nc2} and \ref{fig:CBM_affine_nc5}, multiple comments can be done. First, as expected, a nominal controller $\mathbf K_N$, synthesized on the mean value of the parameter (black dashed curves) cannot perform well over the entire range of parametric variation while the parametric ones obtained in linear fractional form does $\mathbf K_{LF}$ (rational: red solid curves and affine: blue sh dotted). Then, by comparing the two obtained $\mathbf K_{LF}$ controller performances, both provide a performance level below the one obtained during the synthesis (see $\gamma$ values in the legend), which confirms the effectiveness and consistency of the proposed approach. Interestingly, on Figure \ref{fig:CBM_affine_nc2}, the increase of $n_\Delta$ do not necessarily lead to a better attenuation, especially in the rational case (since it adds a lot of variables).
This not the case on Figure \ref{fig:CBM_affine_nc5}. This first experiments shows how easy it is to construct a parametric controller that performs way better than an un-parametric one.

\subsection{Parametric control performances}

In the second application, we consider the invariant Los Angeles Hospital building model extracted from COMPleib \cite{Compleib}. It is a single input single output model of dimension $n=48$ representing the oscillation from a ground to the top of the building. For the purpose of our study, we first have normalized the system so that the $\Hinf$ norm of the open-loop model is equal to one. Then, the generalized plant has been constructed to that the objective of the controller is to attenuate the first amplification peak (around $w_\infty\approx5.2rad/s^{-1}$). Here we aim at finding a family of controller, parametrized along $\pv\in[0.5~1.5]$, an exogenous parameter, that achieve a varying attenuation level on this peak. This attenuation level is classically measured through the $\Htwo$ norm of the transfer function from $\mathbf w(t)$ to $\mathbf z(t)$. To do so, the following generalized plant has been constructed (with varying performances):
\begin{eq}
\left\{
\begin{array}{rcl}
\dx(t) &=& \A \x(t) + \B \mathbf w(t) + \B \u(t)  \\
\mathbf z(t)&=& \C(\pv) \x(t) \\
\y(t)&=& \C\x(t)  
\end{array} \right.  .
\label{eq:geneLAH}
\end{eq}
In addition we impose the following controller structure ($w_m=w_\infty$, $\alpha=10$ and $m=0.1$),
\begin{eq}
Wk = 1/\pv\frac{s^2/(\alpha w_m)^2 + 2ms/w_m+\alpha^{-2}}{s^2/w_m^2+2ms/w_m+1}.
\end{eq}

\begin{figure}
\centering
\includegraphics[width=\columnwidth]{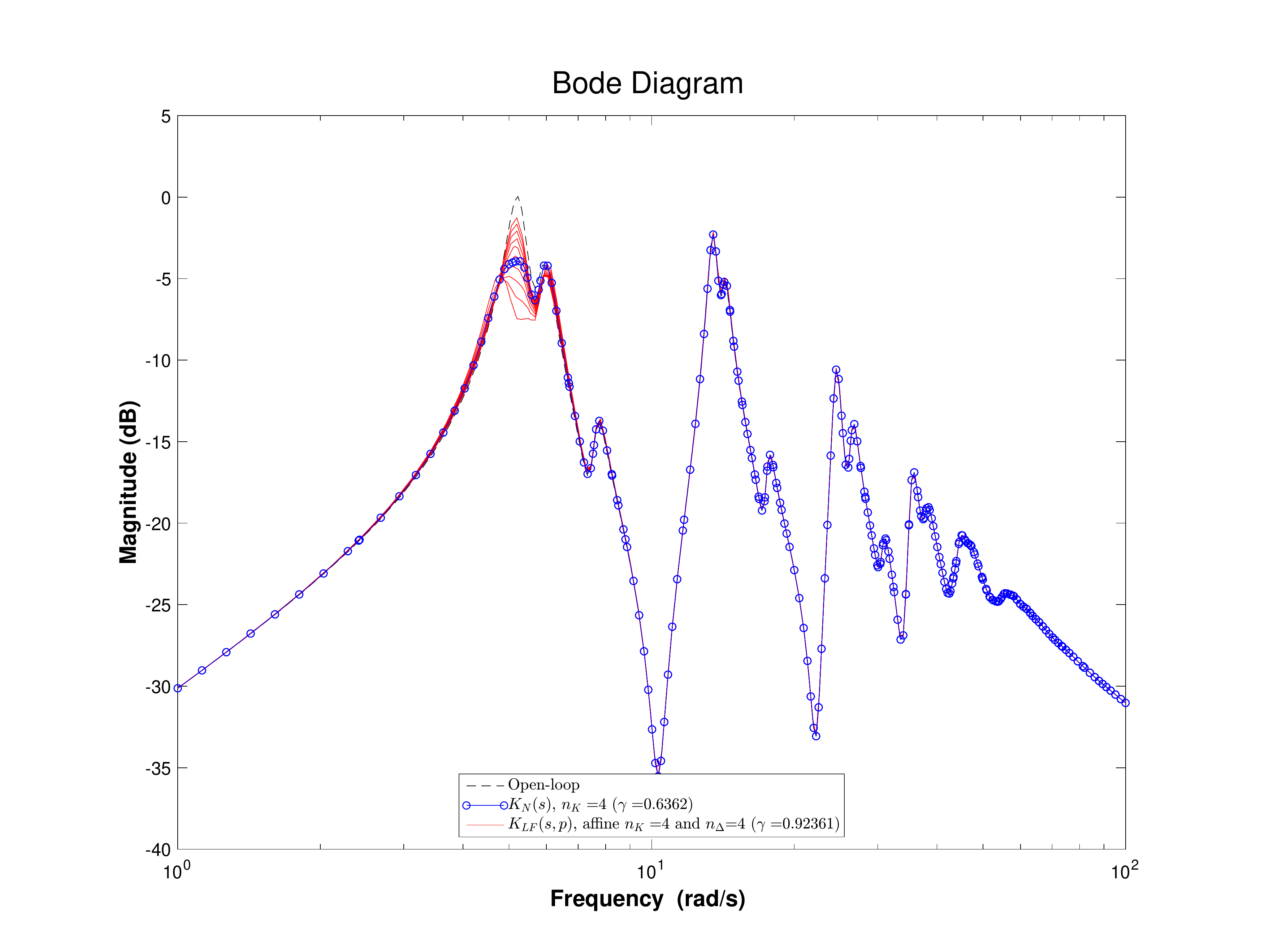}
\caption{Bode diagram of the open-loop (dashed black), the closed-loop obtained with the nominal control $\mathbf K_N(s)$ (obtained for $\pv=1$, blue rounded), and the parametric controller $\mathbf K_{LF}(s,\pv)$ (solid red lines).}
\label{fig:LAH_nc4Bode}
\end{figure}

Figure \ref{fig:LAH_nc4Bode} provides the obtained frequency responses, comparing the open-loop (black dashed), the closed-loop obtained with the nominal control $\mathbf K_N(s)$ (obtained for $\pv=1$, blue rounded), and the parametric controller $\mathbf K_{LF}(s,\pv)$ (solid red lines).

\begin{figure}
\centering
\includegraphics[width=\columnwidth]{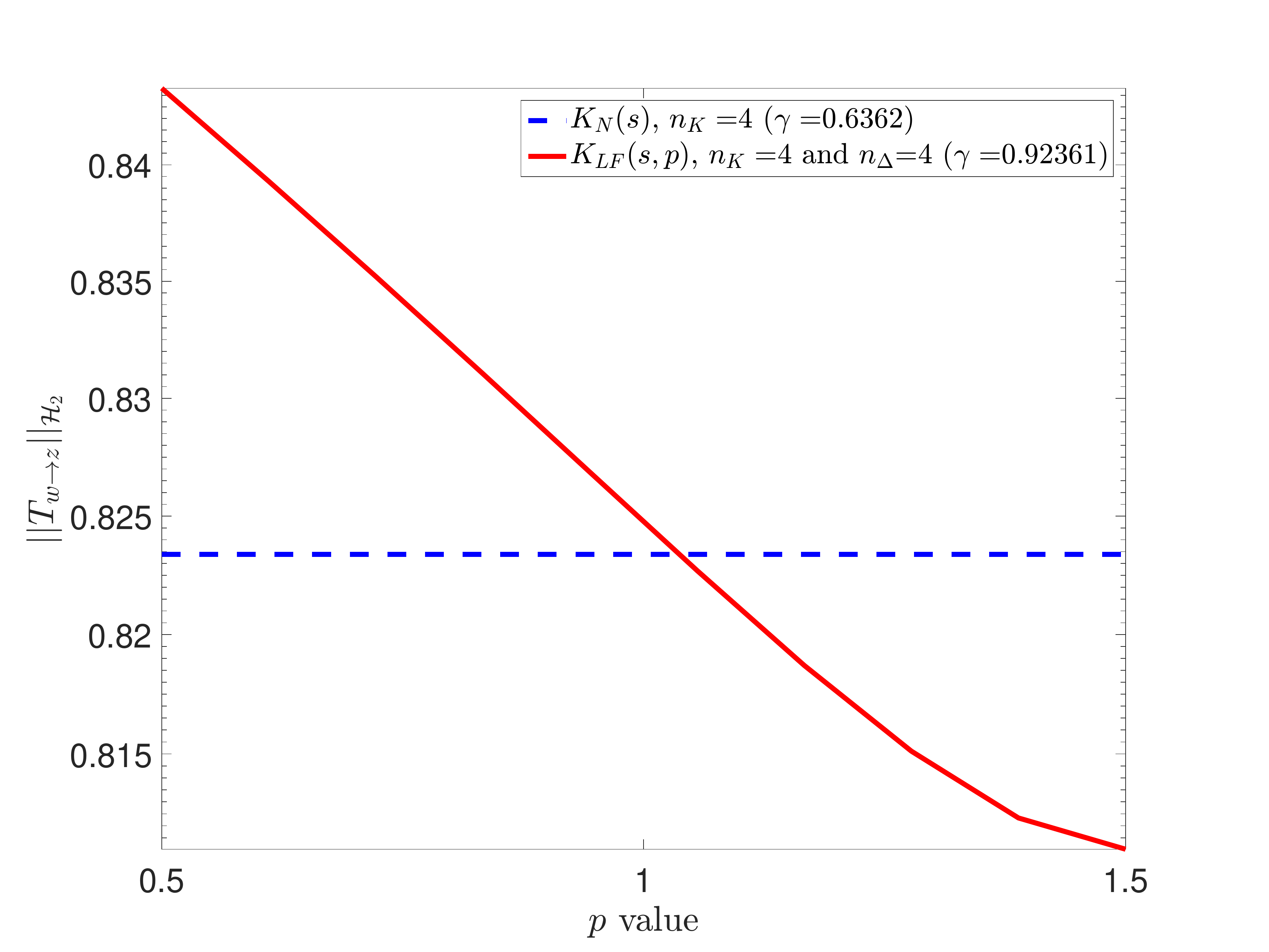}
\caption{$\Htwo$ performance evaluation function of the $\pv$ tuning parameter. Red solid line: looped systems with $\mathbf K_{LF}(s,\pv)$, and dashed blue for $\mathbf K_{N}(s)$.}
\label{fig:LAH_nc4}
\end{figure}

It appears that the control law well reduces the amplification in the region of the first peak with both controllers, as expected by the designer. More interestingly, the parametric controller set $\mathbf K_{LF}(s,\pv)$ shows to provide varying attenuation level performances as $\pv$ varies. this property is highlighted on Figure \ref{fig:LAH_nc4} which illustrates the $\Htwo$ attenuation performances as $\pv$ vary from its minimal to maximal value. The bigger $\pv$ is, the better the attenuation is. One of the very nice property is that for the value of $\pv=1$, $\mathbf K_{LF}(s,1)$, provides similar performances than the un-parametrized control $\mathbf K_{N}(s)$. 

This kind of property, is, in practice, very important if one aims at implementing a controller and try the performances on the real system. As a matter of consequence, the proposed parametric control design show to very well behave, even on quite complex set-up.

\section{Conclusions and perspectives}
\label{section-conclusions}

In this paper, a new simple but effective approach to design low order multiple input multiple output parametric linear fractional controller achieving $\mathcal H_\infty$ performances, has been introduced. The proposed framework is based on the recent developments in $\Hinf$ oriented optimization, which are made available in \cite{Apkarian:2006}. The pivotal idea is based on the specific structure of the control operator, \ie the fractional representation. To the author's feeling, this simple structure, linked with dedicated optimization tools, makes this approach both simple and mathematically well posed, and  stands as a nice solution for many practitioners faced to parametric models and controller synthesis.
Obviously, the results in this paper are not restricted single parameter dependency but yet, it is to be kept in mind that extension to multiple parameters will require dedicated attention due to the complexity increase in the optimization and in the selection of the $n_\Delta$ dimensions. Nevertheless, the approach quickness should be exploited in further developments. Following previous experimental results discussed in \cite{MeyerMOVIC:2016,PoussotCST:2017,MeyerIFASD:2017}, on-going work will implement this strategy on a real aircraft during flight experiments.


\end{document}